# FRACTIONAL TERM STRUCTURE MODELS: NO-ARBITRAGE AND CONSISTENCY

By Alberto Ohashi[1]

*Ibmec-São Paulo and Universidade Estadual de Campinas*

In this work we introduce Heath–Jarrow–Morton (HJM) interest rate models driven by fractional Brownian motions. By using support arguments we prove that the resulting model is arbitrage free under proportional transaction costs in the same spirit of Guasoni [*Math. Finance* **16** (2006) 569–582]. In particular, we obtain a drift condition which is similar in nature to the classical HJM no-arbitrage drift restriction.

The second part of this paper deals with consistency problems related to the fractional HJM dynamics. We give a fairly complete characterization of finite-dimensional invariant manifolds for HJM models with fractional Brownian motion by means of Nagumo-type conditions. As an application, we investigate consistency of Nelson–Siegel family with respect to Ho–Lee and Hull–White models. It turns out that similar to the Brownian case such a family does not go well with the fractional HJM dynamics with deterministic volatility. In fact, there is no nontrivial fractional interest rate model consistent with the Nelson–Siegel family.

**1. Introduction.** Financial models driven by semimartingales and Markov noises have been intensively studied over the last years by many authors. In general, absence of arbitrage is the basic equilibrium condition which fulfills the minimum requirement for any sensible pricing model. On the other hand, empirical studies propose models which are not consistent with this basic assumption. In particular, some evidence of nontrivial long-memory behavior in bond markets has been recently suggested by many authors (see, e.g., [21] and other references therein). In most cases, the presence of long-range dependence in short-rate interest rates seems to be common and it is originated by the fundamentals of the economy. In this regard, it is important to

Received February 2008; revised July 2008.
[1]Supported by FAPESP Grant 05/57064-4.

study bond markets with extrinsic memory driven by non-Markovian noises which allow nontrivial long-range dependence over time.

Recall that in the classical Musiela parametrization the forward rate $r_t$ satisfies a stochastic partial differential equation (henceforth abbreviated by SPDE) of the following form

$$(1.1) \quad dr_t(x) = \left(\frac{\partial}{\partial x} r_t(x) + \alpha_{\mathrm{HJM}}(t, r_t(x))\right) dt + \sum_{j \geq 1} \sigma^j(t, r_t(x)) \, dB_t^j,$$

where $\alpha_{\mathrm{HJM}}$ is the so-called Heath–Jarrow–Morton (henceforth abbreviated by HJM) drift condition which is completely determined by the volatilities $(\sigma^j)_{j \geq 1}$ under a risk-neutral measure, $(B^j)_{j \geq 1}$ is a sequence of stochastic noises and $x$ is the time to maturity. The forward rate $r_t$ is considered as a Hilbert space-valued stochastic process. Due to this infinite-dimensional intrinsic nature, it is important to understand the relation between forward curves $x \mapsto r_t(x)$ at time $t > 0$ and finite-dimensional parameterized families of smooth forward curves, frequently used in estimating the term structure of interest rates (e.g., Nelson–Siegel and Svensson families).

Originally proposed by Björk and Christensen [3] and recently studied by Filipović and Teichmann [10, 11], the so-called *consistency problems* refers to the characterization and existence of finite-dimensional invariant manifolds with respect to $t \mapsto r_t$. In fact, the stochastic invariance is essentially equivalent to a deterministic tangency condition on the coefficients $\frac{\partial}{\partial x}, \alpha_{\mathrm{HJM}}, \sigma$ in (1.1). In particular, if the short-rate of interest rates exhibit long-range dependence then standard statistical procedures may be misspecified, since in this case the classical HJM no-arbitrage drift restriction may not be the correct one. Moreover, by fixing (nonsemimartingale) long-memory stochastic noises $(B_j)_{j \geq 1}$ in (1.1), one has to obtain new tangency conditions on the coefficients of (1.1) to get appropriate arbitrage-free invariant parameterized families of smooth forward curves. This is the program that we start to carry out in this work.

We have chosen the driving noise in (1.1) given by the fractional Brownian motion (henceforth abbreviated by fBm) with Hurst parameter $H \in (1/2, 1)$. For many reasons (see, e.g., [26]), the fBm appears naturally as the canonical process with nontrivial time correlations inserting memory into system under consideration. Indeed, the main difficulty in dealing with fBm is the fact that such process is a semimartingale if and only if it is a standard Brownian motion ($H = 1/2$). This lack of semimartingale property immediately implies that fBm allows arbitrage opportunities (for any $H \neq 1/2$) in the absence of transaction costs. In particular, Gapeev [12] obtains an explicit arbitrage strategy via pathwise integrals in a frictionless bond market driven by a fBm with $H > 1/2$.

The main goal of this work is to introduce arbitrage-free HJM interest rate models driven by a fBm under arbitrary small proportional transaction costs in the bond market. In this paper, the forward rate is considered as the solution of a SPDE (in Skorohod sense) of type (1.1) under the Musiela parametrization. In this work, we only treat the case of deterministic volatilities, leaving open the general stochastic volatility case for future research.

Under deterministic volatility assumption, we obtain a drift condition which is similar in nature to the classical HJM no-arbitrage drift restriction. Although such condition is not sufficient to ensure no-arbitrage in the market, when combined with an additional mild condition on the volatilities it results in absence of arbitrage in the same spirit of the works [14, 15], where the support of the driving noise plays a key rule in the no-arbitrage characterization for markets with transactions costs.

In the second part of this paper, we characterize finite-dimensional invariant submanifolds for HJM models driven by fBm by means of Nagumo-type conditions. Such characterization is the key ingredient to tackle the consistency problems related to the model. As an application of these abstract results, we investigate consistency of the Nelson–Siegel family with respect to Ho–Lee and Hull–White models driven by fBm. Similar to the Brownian case, such a family is not consistent with respect to these models. In general, we arrive at the same classical result of the Brownian case: no nontrivial interest rate model with deterministic volatility structure is consistent with Nelson–Siegel family.

This work is organized as follows. In Section 2, we define the basic SPDE which describes the term structure of the interest rates. In Section 3, we give some general results regarding portfolios and absence of arbitrage in fractional bond markets. In Section 4, we characterize finite-dimensional invariant forward manifolds with respect to HJM models driven by fBm. In Section 5, we examine consistency of the Nelson–Siegel family with respect to concrete interest rate models. The Appendix contains some technical results with respect to the integrals which appear in the self-financing trading strategies introduced in Section 3.

**2. Fractional term structure equations.** Throughout this paper we are given a $d$-dimensional fBm $\beta = (\beta^j)_{j=1}^d$ with parameter $1/2 < H < 1$ on a stochastic basis $(\Omega, (\mathcal{F}_t)_{t \geq 0}, \mathbb{P})$ satisfying the usual conditions. In other words, it is a centered $d$-dimensional Gaussian process with continuous sample paths, $\beta_0 = 0$ and covariance

$$\mathbb{E}(\beta_t^i - \beta_s^i)(\beta_t^j - \beta_s^j) = \delta_{ij}|t-s|^{2H-2}$$

for $t, s \in \mathbb{R}_+$ and $i, j = 1, \ldots, d$. For a detailed discussion on the stochastic analysis of the fBm, the reader may refer to [18, 26].

We denote by $\mathcal{C}_\mathbf{T}$ the space of real-valued continuous functions on a metric space $\mathbf{T}$. In what follows, we consider the following subset of $\mathbb{R}^2$

$$\Delta^2 := \{(t, T) \in \mathbb{R}^2 | 0 \leq t \leq T < \infty\}.$$

Let us consider a term structure of bond prices $\{P(t,T); (t,T) \in \Delta^2\}$ where $P(t,T)$ is the price of a zero coupon bond at time $t$ maturing at time $T$. We assume the usual normalization condition

$$P(t,t) = 1 \qquad \forall t > 0,$$

and $P(t,T)$ is a.s. continuously differentiable in the variable $T$. In this way, we introduce the term structure of interest rates $\{f(t,T); (t,T) \in \Delta^2\}$ given by the following relation:

$$(2.1) \qquad P(t,T) = \exp\left(-\int_t^T f(t,u)\,du\right); \qquad (t,T) \in \Delta^2,$$

for some a.s locally integrable function $f:[t,+\infty) \to \mathbb{R}$, the time $t$ forward curve. In this paper, we adopt the Heath–Jarrow–Morton framework [17] in the fBm setting in terms of the Musiela parametrization where $r_t(x) := f(t,t+x)$ for $(t,x) \in \mathbb{R}_+^2$ and $x = T - t$ is the time to maturity.

In particular, we seek the forward curve $x \mapsto r_t(x)$ as a Hilbert space-valued stochastic process described by a linear SPDE

$$(2.2) \qquad dr_t = (Ar_t + \alpha_t)\,dt + \sum_{i=1}^d \sigma_t^i\,d\beta_t^i, \qquad r_0(\cdot) = \xi \in E$$

in a separable Hilbert space $E$ to be defined. The first-order derivative operator $A := \frac{d}{dx}$ is the infinitesimal generator of the right-shift family of operators $\{S(t); t \geq 0\}$ acting on $E$. Of course, the SPDE (2.2) must be interpreted in the integral form. Moreover, in this work the stochastic integral is considered in the Skorohod sense [9] as a Paley–Wiener integral.

2.1. *The specification of the model.* In this section, we specify a term structure model driven by a $d$-dimensional fBm. Let us assume for the moment that the forward rate is given by the following system of stochastic differential equations

$$(2.3) \qquad f(t,T) = f(0,T) + \int_0^t \alpha(s,T)\,ds + \sum_{i=1}^d \int_0^t \sigma^i(s,T)\,d\beta_s^i.$$

From now on the coefficients $(\sigma^1, \ldots, \sigma^d)$ and $\alpha$ are deterministic functions. Equation (2.3) is well defined if for each $i = 1, \ldots, d$

$$\int_0^T |\alpha(s,T)|\,ds + \int_0^T \int_0^T |\sigma^i(s,T)||\sigma^i(t,T)|\phi_H(t-s)\,ds\,dt < \infty$$

for all $0 < T < \infty$, where $\phi_H(u) := H(2H-1)|u|^{2H-2}$, $u \in \mathbb{R}$.

Let $\{S(t); t \geq 0\}$ be the semigroup of right shifts defined by $S(t)g(x) := g(t+x)$ for any function $g : \mathbb{R}_+ \to \mathbb{R}$. Fix $(t,x) \in \mathbb{R}_+^2$. Then (2.3) can be written as

$$f(t, t+x) = S(t)f(0,x) + \int_0^t S(t-s)\alpha(s, s+x)\, ds$$
(2.4)
$$+ \sum_{i=1}^d \int_0^t S(t-s)\sigma^i(s, s+x)\, d\beta_s^i.$$

In (2.4) we deal with the parametrization $T = t + x$. The operator $S(t)$ acts on $f(0,x)$, $\alpha(s, x+s)$ and $\sigma^j(s, x+s)$ as functions of $x$. By setting

$$r_t(x) := f(t, t+x),$$

it follows that

$$P(t,T) = \exp\left\{-\int_0^{T-t} r_t(x)\, dx\right\}; \qquad (t,T) \in \Delta^2.$$

We can work out in an axiomatic way the minimal requirements on a Hilbert space $E$ such that (2.4) can be given a meaning when

(2.5) $\qquad r_t(\cdot) = f(t, t+\cdot); \qquad t \in \mathbb{R}_+$

is considered as the mild solution of (2.2). The strategy in finding a suitable state space $E$ follows very similar to Filipović [10]; see Section 4.2 and Theorem 5.1.1. So we omit the details and the reader may refer to this work.

REMARK 2.1. Recall that the minimal requirements on the state space $E$ are the following ones: the right-shift semigroup is a $C_0$-semigroup on $E$ and the evaluation mapping $\langle \delta_x, h \rangle = h(x)$ is a bounded linear functional on $E$. We then choose the state space $E$ as defined in [10], Section 5.

In the remainder of this paper, $U$ is a $d$-dimensional vector space with an orthonormal basis $(e_i)_{i=1}^d$ and $\mathcal{L}(U, E)$ is the space of bounded linear operators from $U$ into $E$ with the usual norm $\|\cdot\|$. We make use of the following notation: we set $\alpha_t(\cdot) := \alpha(t, t+\cdot)$ and $\sigma = (\sigma^j)_{j=1}^d$, where

$$\sigma_t^j(\cdot) := \sigma_t e_j(\cdot) := \sigma^j(t, t+\cdot); \qquad j = 1, \ldots, d.$$

In this paper, we are interested in Gaussian interest rate models where we assume that the coefficients $\alpha : \mathbb{R}_+ \to E$ and $\sigma : \mathbb{R}_+ \to \mathcal{L}(U, E)$ satisfy the following set of assumptions:

(2.6) $\qquad \int_0^T \|\alpha_s\|_E\, ds + \int_0^T \|\sigma_s\|^2\, ds < \infty \qquad$ for every $0 < T < \infty.$

To ensure existence of a continuous version for the mild solution of (2.2) we assume there exists $\gamma \in (0, 1/2)$ such that

$$\int_0^T \int_0^T u^{-\gamma} v^{-\gamma} \|S(u)\sigma_u\| \|S(v)\sigma_v\| \phi_H(u-v) \, du \, dv < \infty$$
(2.7)
$$\text{for every } 0 < T < \infty.$$

In order to get a well-defined expression for the bond prices $\{P(t,T); (t,T) \in \Delta^2\}$ we also assume the following growth conditions:

$$\int_{[0,T]^4} \|\sigma_u(s)\|_{\mathbb{R}^d} \|\sigma_v(r)\|_{\mathbb{R}^d} \phi_H(u-v) \, du \, dv \, ds \, dr < \infty$$
(2.8)
$$\text{for every } 0 < T < \infty;$$

$$\int_{[0,T]^3} \|\sigma_u(t)\|_{\mathbb{R}^d} \|\sigma_v(t)\|_{\mathbb{R}^d} \phi_H(u-v) \, dv \, du \, dt < \infty$$
(2.9)
$$\text{for every } 0 < T < \infty.$$

One should note that (2.6) yields $\int_0^T \|S(t)\sigma_t\|^2 \, dt < \infty$ for every $0 < T < \infty$, and therefore the stochastic convolution is a well-defined $E$-valued Gaussian process given by

$$\sum_{j=1}^d \int_0^t S(t-s) \sigma_s^j \, d\beta_s^j, \qquad t > 0.$$

We assume the existence of a traded asset that pays interest. In other words, the unit of money invested at time zero in this asset gives at time $t$ the amount

$$S_0(t) := \exp\left\{\int_0^t r_s(0) \, ds\right\},$$

where $r_t(0) = f(t,t)$ for $t > 0$. By considering $S_0$ as a *numéraire*, the discounted prices are then expressed by

(2.10) $$Z_t(T) := \frac{P(t,T)}{S_0(t)}, \qquad (t,T) \in \Delta^2.$$

In the sequel, we make use of the following notation: if $v : [0,T] \times \mathbb{R}_+ \to \mathbb{R}$ is locally integrable in $\mathbb{R}_+$ then $\mathcal{I}_v(s,T) := \int_0^{T-s} v_s(x) \, dx$. Under the above assumptions one can easily show the following lemma.

LEMMA 2.1. *Assume that the coefficients $\alpha$ and $\sigma$ satisfy the assumptions (2.6), (2.7), (2.8) and (2.9). Then the forward rate $r_t$ is the continuous* *mild solution of (2.2). Moreover, the term structure of bond prices is given by the continuous process*

$$
\begin{aligned}
P(t,T) = P(0,T) \\
\times \exp\left\{\int_0^t [r_s(0) - \mathcal{I}_\alpha(s,T)]\,ds + \sum_{j=1}^d \int_0^t -\mathcal{I}_{\sigma^j}(s,T)\,d\beta_s^j\right\}
\end{aligned}
\tag{2.11}
$$

*for $(t,T) \in \Delta^2$.*

PROOF. Equation (2.11) is a straightforward application of stochastic Fubini theorem (see [20]) in the fBm setting by using conditions (2.8) and (2.9). Conditions (2.6) and (2.7) allow the existence of a continuous mild solution of (2.2) as in [9]. □

**3. The bond market: Portfolios and no-arbitrage.** In this section, we discuss the basic setting for a fractional term structure of bond prices with transaction costs. It is well known that not every choice of stochastic integral makes sense in Finance (see, e.g., [5]). Therefore, at this point it is necessary to discuss if our choice of stochastic integral presents conceptual problems. We recall that integrals with respect to $\mathcal{C}_{\mathbb{R}_+}$-valued processes play a key rule in the bond market theory. The integration theory for Banach space-valued semimartingales can be developed in many ways (see, e.g., [4, 6, 8]). In general, the fundamental idea is to make use of duality arguments between a "state variable" $S_t$ (which represents a discounted price curve at time $t$ in $\mathcal{C}_\mathbf{T}$) and a dual $\mathcal{C}_\mathbf{T}^*$-valued process $\mu$ which represents a portfolio strategy. We stress here that this procedure is fully related to the intrinsic mechanism of a bond market and *not* to a given stochastic dynamics on $S$. In this way, we may adopt the same procedure in the nonsemimartingale case.

Since we are dealing with a continuous Gaussian HJM model we shall make use of a mild integrability assumption [see (H1)] to define the correspondent wealth processes without any use of the *fine* structure of the Skorohod integrals which drive $Z$. In particular, by a standard limit procedure we avoid the Wick products in the definition of self-financing portfolios (see, e.g., [5] for a discussion on stock markets). For us, the most important property is an integration by parts formula (see Proposition A.1) in the same spirit of the works [14, 15].

We now introduce the notions of admissible self-financing portfolios in our context. Let us denote by $\mathcal{M}_{T^*}$ the space of (finite) signed measures on $[0,T^*]$ endowed with the total variation norm $\|\cdot\|_{\mathrm{TV}}$. Let $\mu$ be a measure-valued elementary process of the form

$$
\mu_t(\omega,\cdot) := \sum_{i=0}^{N-1} \chi_{F_i \times (t_i, t_{i+1}]}(\omega, t) m_i(\cdot),
\tag{3.1}
$$

where $m_i \in \mathcal{M}_{T^*}$, $0 = t_0 < \cdots < T_N \leq T^*$ and $F_i \in \mathcal{F}_{t_i}$.

We denote by $\mathcal{S}_b$ the set of elementary processes of the form (3.1) endowed with the following norm:

$$\|\mu\|_V^2 := \mathbb{E} \sup_{0 \leq t \leq T^*} \|\mu_t\|_{\mathrm{TV}}^2. \tag{3.2}$$

From now on all economic activity will be assumed to take place on the bounded set $[0, T^*]^2$. So we assume that $Z_t(T) = 0$ if $(t, T) \notin [0, T^*]^2$. Under the hypotheses (2.6), (2.7), (2.8) and (2.9), the discounted price process $Z_t(T)$ satisfies the following condition:

(H1) $\{Z_t(T); (t, T) \in [0, T^*]^2\}$ is a jointly continuous real-valued stochastic process such that

$$\mathbb{E} \sup_{(t,T) \in [0,T^*]^2} |Z_t(T)|^2 < \infty.$$

If $\mu \in \mathcal{S}_b$ is given by (3.1) then we define

$$\int_0^t \mu_s \, dZ_s := \sum_{i=0}^{N-1} \chi_{F_i} (Z_{t_{i+1} \wedge t} - Z_{t_i \wedge t}) m_i, \tag{3.3}$$

where $Z_{t_i} m_{t_i}$ is the usual dual action. By Hölder's inequality it follows that

$$\mathbb{E} \sup_{0 \leq t \leq T^*} \left| \int_0^t \mu_s \, dZ_s \right| \leq \|\mu\|_V \mathbb{E}^{1/2} \sup_{0 \leq s, t \leq T^*} \|Z_s - Z_t\|_\infty^2 < \infty, \tag{3.4}$$

where $\|\cdot\|_\infty$ denotes the usual (uniform topology) norm on the space of real-valued bounded functions defined on $[0, T^*]$. Let $V$ be the completion of $\mathcal{S}_b$ with respect to (3.2). By the estimate (3.4) and the definition of $V$ we may easily define $\int_0^\cdot \mu_s \, dZ_s$ for every $\mu \in V$.

In the sequel, we denote $\mathcal{P}_{T^*}$ the set of all partitions of $[0, T^*]$. We also need the following assumption:

(H2) $\Pi_{T^*}(\mu) := \sup_{\pi \in \mathcal{P}_{T^*}} \sum_{t_i \in \pi} \|\mu_{t_{i+1}} - \mu_{t_i}\|_{\mathrm{TV}}$ is square integrable.

By taking into account proportional transaction costs in the bond market, the liquidation value of a portfolio with zero initial capital is

$$V_t^k(\mu) := \sum_{t_i < t} \chi_{F_i} (Z_{t_{i+1} \wedge t} - Z_{t_i \wedge t}) m_i$$

$$- k \sum_{t_i < t} Z_{t_i} |\mu_{t_{i+1}} - \mu_{t_i}| - k Z_t |\mu_t|,$$

where $k$ is an arbitrary positive number and $|\cdot|$ denotes the total variation measure. The first term accounts for the capital gain of holding an elementary strategy $\mu$ of the form (3.1) (without transaction costs) during the interval $[0, t]$. The second and third term account for the transaction costs incurred in various transactions and the eventual liquidation value of the portfolio, respectively.

By passing from a finite number of transactions to continuous trading one can easily show that if $\mu \in V$ satisfies assumption (H2) then the above quantities converge to the following:

$$(3.5) \qquad V_t^k(\mu) := \int_0^t \mu_s \, dZ_s - k \int_0^t Z_s \, d|\mu_s| - kZ_t|\mu_t|.$$

In other words, under assumptions (H1) and (H2) it follows that any wealth process is the limit of elementary strategies in the sense that

$$\lim_{n \to \infty} \mathbb{E} \sup_{0 \le t \le T^*} |V_t^k(\mu^n) - V_t^k(\mu)| = 0; \qquad k > 0,$$

where $\mu^n$ is a sequence of elementary strategies such that $\mu^n \to \mu$ in V as $n \to \infty$. See the Appendix for more details, including the definition of the second integral in (3.5).

REMARK 3.1. One may also introduce another approach to define the integral for measure-valued integrands; the integration theory is reduced via Fubini theorems where the stochastic differential is interpreted in the Stratonovich sense [24]. In this case, $\int_0^t \mu_s \, dX_s$ is replaced by

$$(\mu \circ X)_t = \int_0^t \mu_s \xi_s \, ds + \sum_{i=1}^d \int_0^t \mu_s \rho_s^i \, d\beta_s^i,$$

where $dX_t(T) = \xi_t(T) \, dt + \sum_{i=1}^d \rho_t^i(T) \, d\beta_s^i$ is a SDE satisfying (H1). One can easily show that under additional standard Hölder assumptions on the paths of $\mu \in V$ and $\sigma$ (see, e.g., [2, 24]), we do have $\int_0^t \mu_s \, dZ_s = (\mu \circ Z)_t$ for $0 \le t \le T$.

Now we are able to introduce the following notions:

DEFINITION 3.1. We say that $\mu \in V$ is an *admissible trading strategy* if it satisfies (H2), it is (weakly) $\mathcal{F}_t$-adapted and there exists a constant $M > 0$ such that $V_t^k(\mu) \ge -M$ a.s. for every $t \le T^*$. An admissible trading strategy is an *arbitrage opportunity with transaction costs* $k > 0$ on $[0, T^*]$, if $V_{T^*}^k(\mu) \ge 0$ a.s. and $\mathbb{P}\{V_{T^*}^k(\mu) > 0\} > 0$. Therefore, the bond market is *k-arbitrage free* on $[0, T^*]$ with transaction costs $k$ if for every admissible strategy $\mu$, $V_{T^*}^k(\mu) \ge 0$ a.s. only if $V_{T^*}^k(\mu) = 0$ a.s.

REMARK 3.2. Since the main dynamics takes place on $\Delta^2$ we do assume that the support of $\mu_s$ is concentrated on $[s, +\infty)$ for every admissible strategy $\mu$.

It is straightforward to prove the following result in the same spirit of Guasoni ([14], Proposition 2.1) by using the integration by parts formula (A.1). So we omit the details.

PROPOSITION 3.1. *Let us fix $k > 0$. If for every $(\mathcal{F}_t)_{t \geq 0}$-stopping time $\tau$ such that $\mathbb{P}\{\tau < T^*\} > 0$ we have*

$$(3.6) \qquad \mathbb{P}\left\{\sup_{\tau \leq t \leq T \leq T^*}\left|\frac{Z_\tau(\tau)}{Z_t(T)} - 1\right| < k, \tau < T^*\right\} > 0,$$

*then the bond market is arbitrage free on $[0, T^*]$ with transaction costs $k$.*

One can also show by using similar arguments from Guasoni, Rásonyi and Schachermayer [15] that the $k$-arbitrage-free property in Definition 3.1 is essentially equivalent to the existence of a $k$-consistent price system. Thus (3.6) is also a sufficient condition for it. In fact, a sufficient condition for no-arbitrage is the conditional full support property for $Z$ which is equivalent to full support only if $Z$ is Markovian. Hence in the non-Markovian setting it is more natural to find mild conditions on the volatility in such way that $\log Z$ has only full support. See Lemmas 3.1 and 3.2 for details.

3.1. *Absence of arbitrage.* We prove that under suitable conditions on the volatility $\sigma = (\sigma^j)_{j=1}^d$, the bond market model is $k$-arbitrage free for every $k > 0$. The main ingredient in the no-arbitrage argument consists in the full support property on $\mathcal{C}_{\Delta_{T^*}^2}$, where $\Delta_{T^*}^2 := \{(t, T); 0 \leq t \leq T \leq T^*\}$. This property together with a suitable choice on the drift will result in $k$-no-arbitrage for every $k > 0$. Recall that if $\mathcal{X}$ is a Polish space then a random element $\xi : \Omega \to \mathcal{X}$ has $\mathbb{P}$-full support when $\mathbb{P}_\xi := \mathbb{P} \circ \xi^{-1}(\mathcal{U}) > 0$ for every nonempty open set $\mathcal{U}$ in $\mathcal{X}$.

LEMMA 3.1. *Let $\mathbb{Y} : \Omega \to \mathcal{C}_{\Delta_{T^*}^2}$ be a measurable map such that $\mathbb{X} := \log \mathbb{Y}$ has $\mathbb{P}$-full support. Then $\mathbb{Y}$ satisfies assumption in Proposition 3.1.*

PROOF. Given $\varepsilon > 0$ and $\tau$ a $\mathcal{F}_t$-stopping time such that $\mathbb{P}\{\tau < T^*\} > 0$, it is sufficient to check that

$$\mathbb{P}\left\{\sup_{\tau \leq t \leq T \leq T^*}|\mathbb{X}(t, T) - \mathbb{X}(\tau, \tau)| < \varepsilon, \tau < T^*\right\} > 0.$$

If $p \in \mathcal{C}_{\Delta_{T^*}^2}$ then triangle inequality yields

$$\left\{\sup_{(t,T) \in \Delta_{T^*}^2}|\mathbb{X}(t, T) - p(t, T)| < \varepsilon/2, \tau < T^*\right\}$$

$$\subset \left\{\sup_{\tau \leq t \leq T \leq T^*}|\mathbb{X}(t, T) - \mathbb{X}(\tau, \tau)| < \varepsilon, \tau < T^*\right\}.$$

Let us consider $\mathcal{P}$ the set of polynomials $p$ on $\Delta_{T^*}^2$ with rational coefficients such that $p(0, 0) = 0$. We claim that there exists $p \in \mathcal{P}$ such that

$$(3.7) \qquad \mathbb{P}\left\{\sup_{(t,T) \in \Delta_{T^*}^2}|\mathbb{X}(t, T) - p(t, T)| < \varepsilon/2, \tau < T^*\right\} > 0.$$

Suppose that (3.7) is violated for every $p \in \mathcal{P}$. Then we obtain

$$\left\{\sup_{(t,T)\in\Delta^2_{T^*}} |\mathbb{X}(t,T) - p(t,T)| < \varepsilon/2, \tau < T^*\right\}$$
$$\subset \{\tau \geq T^*\} \qquad \mathbb{P}\text{-a.s. } \forall p \in \mathcal{P}.$$

Therefore

$$(3.8) \quad \bigcup_{p\in\mathcal{P}} \left\{\sup_{(t,T)\in\Delta^2_{T^*}} |\mathbb{X}(t,T) - p(t,T)| < \varepsilon/2\right\} \subset \{\tau \geq T^*\} \qquad \mathbb{P}\text{-a.s.}$$

By the density of $\mathcal{P}$ in $\mathcal{C}_{\Delta^2_{T^*}}$ and the full support of $\mathbb{X}$ it follows that

$$\mathbb{P}\left\{\bigcup_{p\in\mathcal{P}} \left\{\sup_{(t,T)\in\Delta^2_{T^*}} |\mathbb{X}(t,T) - p(t,T)| < \varepsilon/2\right\}\right\} = 1$$

and therefore $\mathbb{P}\{\tau < T^*\} = 0$ which is a contradiction. □

REMARK 3.3. Recall that the fBm has $\gamma$-Hölder continuous paths a.s. for any $\gamma < H$. Moreover, one can prove the existence of the fBm Wiener measure on a separable Banach space $\mathbb{W}$ continuously imbedded on the space $\mathcal{C}_{\mathbb{R}_+}$ such that the elements of $\mathbb{W}$ are $\gamma$-Hölder continuous functions on any compact interval. See [16] for the proof of this fact.

The following remark turns out to be very useful for the approach taken in this work.

LEMMA 3.2. Assume that $\mathcal{I}_{\sigma^j}(t,T)$ is $\lambda$-Hölder continuous on $\Delta^2_{T^*}$ for every $j \geq 1$ where $1/2 < \lambda < 1$. Then the process $\sum_{j=1}^{d} \int_0^t \mathcal{I}_{\sigma^j}(s,T) \, d\beta^j_s$ has $\mathbb{P}$-full support on $\mathcal{C}_{\Delta^2_{T^*}}$.

PROOF. Fix $(\xi^j)_{j=1}^d$ a sequence of $\gamma$-Hölder continuous functions on $[0,T^*]$ where $1/2 < \gamma < H$. We recall that if $\mathcal{I}_{\sigma^j}(t,T)$ is $\lambda$-Hölder continuous on $\Delta^2_{T^*}$ then the pathwise Young integral $\int_0^t \mathcal{I}_{\sigma^j}(s,T) \, d\xi^j_s$ is well defined and there exists a constant $C > 0$ which depends only on $T^*$, $\gamma$ and $\lambda$ such that

$$(3.9) \quad \left\|\int_0^\cdot \mathcal{I}_{\sigma^j}(s,\cdot) \, d\xi^j_s\right\|_\gamma \leq C \|\mathcal{I}_{\sigma^j}\|_\lambda \|\xi^j\|_\gamma, \qquad j = 1, \ldots, d,$$

where $\|\cdot\|_\eta$ denotes the usual $\eta$-Hölder norm.

Moreover, the pathwise Young integral coincides with the symmetric integral in Russo and Vallois [24]. Recall that we are assuming that the volatilities are deterministic functions and therefore the Gross–Sobolev derivative of $\mathcal{I}_{\sigma^j}(t,T)$ vanishes for each $j \geq 1$ and $(t,T) \in \Delta^2_{T^*}$. Since the fBm has $\gamma$-Hölder continuous paths a.s., Proposition 3 in [2] implies that the Skorohod integral

$$\int_0^t \mathcal{I}_{\sigma^j}(s,T)\, d\beta^j_s; \qquad j \geq 1,$$

can be interpreted as a pathwise Young integral. By the estimate (3.9) and Remark 3.3 it follows that each $\int_0^\cdot \mathcal{I}_{\sigma^j}(s,\cdot)\, d\beta^j_s$ has $\mathbb{P}$-full support on $\mathcal{C}_{\Delta^2_{T^*}}$. Moreover, since $(\beta^j)_{j\geq 1}$ is a sequence of real-valued independent fBm we then conclude that

$$(t,T) \mapsto \sum_{j=1}^d \int_0^t \mathcal{I}_{\sigma^j}(s,T)\, d\beta^j_s$$

has $\mathbb{P}$-full support as well. $\square$

By Lemma 3.1 and Proposition 3.1 we know that if $\log Z$ has $\mathbb{P}$-full support then the bond market is $k$-arbitrage free for every $k > 0$. Moreover, if the volatility $\sigma = (\sigma^j)_{j=1}^d$ satisfies the assumptions in Lemma 3.2, there are infinitely many choices of $\alpha$ which give the full support property for $\log Z$ and therefore absence of arbitrage in the fractional bond market. But there is a canonical choice for the drift which gives the desirable property (see Definition 3.2, Remark 3.5 and Proposition 3.2): for every $0 < T < \infty$

(3.10) $$\mathbb{E} Z_t(T) = P(0,T), \qquad 0 \leq t \leq T.$$

As a direct consequence of Lemma 2.1 we have the following basic result.

COROLLARY 3.1. *Condition (3.10) holds if and only if the drift $\alpha$ satisfies the following equality*

(3.11)
$$\alpha_t(\cdot) = \sum_{j=1}^d \left\{ \sigma^j_t(\cdot) \int_0^t \mathcal{I}_{\sigma^j}(\theta, \cdot + t)\phi_H(t-\theta)\, d\theta \right. \\ \left. + \int_0^\cdot \sigma^j_t(y)\, dy \int_0^t \sigma^j_\theta(\cdot + t - \theta)\phi_H(t-\theta)\, d\theta \right\}.$$

PROOF. By Lemma 2.1 it follows that

$$Z_t(T) = P(0,T)\exp\left\{ -\int_0^t \mathcal{I}_\alpha(s,T)\, ds - \sum_{j=1}^d \int_0^t \mathcal{I}_{\sigma^j}(s,T)\, d\beta^j_s \right\};$$

$$(t,T) \in \Delta^2.$$

If $0 < T < \infty$, then the Itô formula [2] and standard Fubini theorem yield the following expression:

$$(3.12) \quad y(t,T) = 1 + \sum_{j=1}^{d} \int_0^t y(s,T) \mathcal{I}_{\sigma^j}(s,T) \left[ \int_0^s \mathcal{I}_{\sigma^j}(\theta,T) \phi_H(s-\theta) \, d\theta \right] ds,$$

where $y(t,T) = \mathbb{E} \exp X(t,T)$, $X(t,T) = -\sum_{j=1}^{d} \int_0^t \mathcal{I}_{\sigma^j}(s,T) \, d\beta_s^j$ for $0 \leq t \leq T$. By the variation of constants formula it follows that

$$y(t,T) = \exp\left( \int_0^t e(s,T) \, ds \right),$$

where

$$(3.13) \quad e(t,T) = \sum_{j=1}^{d} \mathcal{I}_{\sigma^j}(t,T) \int_0^t \mathcal{I}_{\sigma^j}(\theta,T) \phi_H(t-\theta) \, d\theta, \qquad 0 \leq t \leq T.$$

Therefore, we arrive at the following conclusion: condition (3.10) holds if, and only if, for every $0 < T < \infty$

$$(3.14) \quad \int_0^{T-t} \alpha_t(y) \, dy = e(t,T), \qquad 0 \leq t \leq T.$$

By differentiating expression (3.14) and making the change of variables $x = T - t$, we see that (3.10) holds if, and only if, $\alpha_t$ satisfies (3.11). □

In view of Corollary 3.1 we now define

$$(3.15) \quad \mathcal{S}_H \sigma_t(\cdot) := \sum_{j=1}^{d} \left\{ \sigma_t^j(\cdot) \int_0^t \mathcal{I}_{\sigma^j}(\theta, \cdot + t) \phi_H(t-\theta) \, d\theta \right. \\ \left. + \int_0^{\cdot} \sigma_t^j(y) \, dy \int_0^t \sigma_\theta^j(\cdot + t - \theta) \phi_H(t-\theta) \, d\theta \right\}$$

and we assume that the volatilities are regular enough in such a way that

$$(3.16) \quad \int_0^T \|\mathcal{S}_H \sigma_t\|_E \, dt < \infty$$

for every $0 < T < \infty$. Indeed, it is not very restrictive to assume that the volatility $\sigma_t$ satisfies such integrability condition on the forward curve space $E$ given in Remark 2.1. See [10], Section 5.2, for more details.

3.2. *Drift condition and quasi-martingale measure.* Similar to the semimartingale case, the measure $\mathbb{P}$ is considered as physical measure. This motivates the following definition.

DEFINITION 3.2. We say that an equivalent probability measure $\mathcal{Q} \sim \mathbb{P}$ is a *quasi-martingale measure* if the discounted bond price process $Z_t(T)$ has $\mathcal{Q}$-constant expectation, that is, for every $0 < T < \infty$,

$$\mathbb{E}_{\mathcal{Q}} Z_t(T) = P(0,T), \qquad 0 \leq t \leq T. \tag{3.17}$$

REMARK 3.4. This notion was already introduced in [27] in a fractional Black–Scholes economy under the name *average risk neutral measure* and also used in [12] in a frictionless fractional bond market in the context of pathwise integration.

REMARK 3.5. One should notice that the change from the actual measure $\mathbb{P}$ to a quasi-martingale measure changes the mean rate of return of the bond but not the volatility. In fact, under a quasi-martingale measure, all financial assets in the bond market have the same expected rate of return, regardless of the "riskiness" of the bond. This is in contrast to the physical measure where more risky assets have a greater expected rate of return than less risky assets.

We now state the main result of this section. Before this, we recall an elementary result concerning a Girsanov theorem in the fBm setting. Without any loss of generality we take $U = \mathbb{R}^d$. Let $\mathbb{H}$ be the Cameron–Martin space associated to the fBm. Recall that $\mathbb{H} = \text{Image}\,\mathcal{K}$ where

$$\mathcal{K}h(t) := \int_0^t K(t,s)h(s)\,ds; \qquad h \in L^2(0,T^*;\mathbb{R}^d) \tag{3.18}$$

is an isomorphism between $L^2(0,T^*;\mathbb{R}^d)$ and $\mathbb{H}$. The kernel $K$ in (3.18) is given by

$$K(t,s) := c_H s^{1/2-H} \int_s^t (u-s)^{H-3/2} u^{H-1/2}\,du \tag{3.19}$$

for some $c_H > 0$ (see, e.g., [2, 18] for more details). The next result is a straightforward consequence of the representation of fBm in terms of the standard Brownian motion.

LEMMA 3.3. *Let $B = (\beta^1, \ldots, \beta^d)$ be a d-dimensional fBm and let $\{\gamma(t); 0 \leq t \leq T^*\}$ be an $\mathbb{R}^d$-valued measurable function such that $\int_0^{T^*} \|\gamma(t)\|_{\mathbb{R}^d}\,dt < \infty$ and $R(\cdot) := \int_0^{\cdot} \gamma(s)\,ds \in \mathbb{H}$. Then $\tilde{B}_t := B_t - \int_0^t \gamma(s)\,ds$ is a d-dimensional $\mathcal{Q}_{T^*}$-fBm on $[0,T^*]$ such that*

$$\frac{d\mathcal{Q}_{T^*}}{d\mathbb{P}} = \mathcal{E}(\mathcal{K}^{-1}R \cdot W)_{T^*}, \tag{3.20}$$

*where*

$$\mathcal{E}(\mathcal{K}^{-1}R \cdot W)_{T^*} := \exp\left[(\mathcal{K}^{-1}R \cdot W)_{T^*} - \frac{1}{2}\int_0^{T^*} \|\mathcal{K}^{-1}R(t)\|_{\mathbb{R}^d}^2 \, dt\right],$$

*and $(\mathcal{K}^{-1}R \cdot W)_{T^*}$ is the usual Itô stochastic integral with respect to the Brownian motion $W$ associated to $B$. In this case, we may write*

$$\tilde{B}_t = \sum_{j=1}^d \tilde{\beta}_t^j e_j,$$

*where $\tilde{\beta}_t^j := \beta_t^j - \int_0^t \gamma_s^j \, ds$ is a $\mathcal{Q}_{T^*}$-real valued independent fBm for each $1 \leq j \leq d$.*

Recall that all economic activity is assumed to take place on the finite horizon $[0, T^*]$. Let us fix $k > 0$ which corresponds to arbitrary small proportional transaction costs in the bond market. The main result of this section is then the following.

THEOREM 3.1. *Assume that the volatility satisfies assumptions in Lemma 3.2 and there exists an $\mathbb{R}^d$-valued measurable function $\gamma_t$ satisfying assumptions in Lemma 3.3 in such way that*

(3.21) $$\sigma_t \gamma_t = \mathcal{S}_H \sigma_t - \alpha_t; \qquad t \geq 0.$$

*Then there exists a quasi-martingale measure for the bond market. In addition, the market is arbitrage free on $[0, T^*]$ with transaction costs $k$.*

PROOF. The forward rate is the continuous mild solution of

$$dr_t = (Ar_t + \alpha_t)\,dt + \sum_{j=1}^d \sigma_t^j \, d\beta_t^j,$$

under the measure $\mathbb{P}$. Assuming assumptions in Lemma 3.3 and (3.21), we may write

$$dr_t = (Ar_t + \mathcal{S}_H \sigma_t)\,dt + \sum_{j=1}^d \sigma_t^j \, d\tilde{\beta}_t^j$$

under the equivalent probability measure $\mathcal{Q}_{T^*}$ with respect to $\mathbb{P}$ given in (3.20). By (3.16) one should notice that the above equation is well defined under $\mathcal{Q}_{T^*}$. By changing the measure $\mathbb{P}$ to $\mathcal{Q}_{T^*}$ in Corollary 3.1 it follows that for every $0 < T < \infty$

$$\mathbb{E}_{\mathcal{Q}_{T^*}} Z_t(T) = P(0, T); \qquad 0 \leq t \leq T,$$

and therefore $\mathcal{Q}_{T^*}$ is a quasi-martingale measure. By Lemma 3.2 the two parameter process $(t,T) \mapsto \sum_{j=1}^{d} \int_0^t \mathcal{I}_{\sigma^j}(s,T) \, d\tilde{\beta}_s^j$ has $\mathcal{Q}_{T^*}$-full support and therefore $\log Z$ has $\mathcal{Q}_{T^*}$-full support as well. By Lemma 3.1 and Proposition 3.1 we conclude the proof. □

REMARK 3.6. One should notice that if there exists a quasi-martingale measure then it should be of the form (3.20), (3.21).

The next result gives an explicit formula for the term structure of bond prices in terms of a conditional expectation. By representing the fBm in terms of a stochastic convolution with respect to a Gaussian martingale $M^j$ (see, e.g., [23]) one can easily obtain an explicit formula for the term structure of bond prices. In the sequel, we denote

$$\theta^j(r,t) := \int_r^t \sigma^j(s,t) s^{H-1/2} (s-r)^{H-3/2} \, ds, \qquad j=1,\ldots,d,$$

for $0 < r < t < \infty$. We also write $[M^j]$ to denote the usual quadratic variation of the martingale $M^j$.

PROPOSITION 3.2. *Assume that $\mathcal{Q}$ is a quasi-martingale measure. Then the bond price can be expressed by*

$$(3.22) \qquad P(t,T) = e^{\xi(t,T)} \mathbb{E}_{\mathcal{Q}} \left[ \exp\left(-\int_t^T r_s(0) \, ds\right) \Big| \mathcal{F}_t \right],$$

*where the kernel $\xi(t,T)$ is given by*

$$\xi(t,T) := -\int_t^T \mathcal{I}_{\mathcal{S}_H \sigma}(s,T) \, ds - \sum_{j=1}^d \int_0^t \mathcal{I}_{\sigma^j}(s,T) \, d\tilde{\beta}_s^j - G(t,T)$$

*and* $G(t,T) := \sum_{j=1}^d \int_0^t \int_r^T \theta^j(r,u) \, du \, dM_r^j - \frac{1}{2} \sum_{j=1}^d \int_t^T (\int_r^T \theta^j(r,u) \, du)^2 \, d[M^j]_r.$

PROOF. Formulas (2.11) and (2.10) yield the following expression:

$$P(t,T) = e^{\xi(t,T)} \mathbb{E}_{\mathcal{Q}} \left[ \exp\left(-\int_t^T r_s(0) \, ds\right) \Big| \mathcal{F}_t \right],$$

where

$$\xi(t,T) = -\int_0^t \mathcal{I}_{\mathcal{S}_H \sigma}(s,T) \, ds - \sum_{j=1}^d \int_0^t \mathcal{I}_{\sigma^j}(s,T) \, d\tilde{\beta}_s^j$$
$$- \ln \mathbb{E}_{\mathcal{Q}} \left[ \exp\left(-\sum_{j=1}^d \int_0^T \mathcal{I}_{\sigma^j}(s,T) \, d\tilde{\beta}_s^j\right) \Big| \mathcal{F}_t \right].$$

We only need to compute the above conditional expectation. By using the representation given in [23] it follows that

$$\int_0^t \sigma^j(s,t)\,d\tilde{\beta}_s^j = \int_0^t \theta^j(r,t)\,dM_r^j.$$

By using conditions (2.6), (2.8) and (2.9) and changing the order of integration, we obtain

$$\int_0^T \mathcal{I}_{\sigma^j}(s,T)\,d\tilde{\beta}_s^j = \int_0^T \int_r^t \theta^j(r,u)\,du\,dM_r^j, \qquad j=1,\ldots,d,$$

and therefore the conditional expectation can be written as

$$\sum_{j=1}^d \int_0^t \int_r^T \theta^j(r,u)\,du\,dM_r^j - \frac{1}{2}\sum_{j=1}^d \int_t^T \left(\int_r^T \theta^j(r,u)\,du\right)^2 d[M^j]_r. \qquad \square$$

REMARK 3.7. A similar pricing formula was obtained in [12], formula (16). One should notice a slight difference between these two formulas due to the fact that the drift computed in [12]; formula (15), which realizes (3.10) is not the correct one. See (3.13) and (3.14) for the proof of this fact.

REMARK 3.8. Besides giving an explicit expression for the term structure of bond prices, formula (3.22) can also be used for pricing contingent claims in a formal way as follows. Let $\mathcal{Q}$ be a quasi-martingale measure and assume that $X \in L^1(\Omega, \mathcal{F}_T, \mathcal{Q}_T)$ is a claim due at time $T$. Then it follows that

$$(3.23) \qquad P(t,T)\mathbb{E}_{\mathcal{Q}_T}[X|\mathcal{F}_t] = S_0(t)\mathbb{E}_{\mathcal{Q}}\left[\frac{X}{S_0(T)}\bigg|\mathcal{F}_t\right] \qquad \text{a.s. } \forall t \in [0,T],$$

where the right-hand side can be formally interpreted as the price of the claim at time $t$ and $\mathcal{Q}_T \sim \mathcal{Q}$ is defined by

$$(3.24) \qquad \frac{d\mathcal{Q}_T}{d\mathcal{Q}} = \frac{P(T,T)}{S_0(T)P(0,T)}.$$

Formula (3.23) is very useful as long as one can compute the distribution of $X$ under $\mathcal{Q}_T$. In the semimartingale case, this is done by means of the martingale property $\mathbb{E}_{\mathcal{Q}}[\frac{d\mathcal{Q}_T}{d\mathcal{Q}}|\mathcal{F}_t] = Z_t(T)/P(0,T)$. In the fractional case, one can express this conditional expectation in terms of a prediction formula in the same spirit of [13].

We now show some examples of familiar short-rate models in the fBm setting as developed in this section under the quasi-martingale $\mathcal{Q}$.

EXAMPLE 1 (Ho–Lee). Let us assume that $d=1$ and $\sigma_t(x) = \sigma$, a constant, for all $(t,x) \in \mathbb{R}_+^2$. Then in this case the model is $k$-arbitrage free for every $k>0$ and the short-rate dynamics under $\mathcal{Q}$ is given by

$$r_t(0) = r_0(t) + \sigma^2 \int_0^t \int_0^s [2t - (s+\theta)]\phi_H(s-\theta)\, d\theta\, ds + \sigma\tilde{\beta}_t,$$

and we recognize this as the Ho–Lee model with a deterministic time-varying drift.

EXAMPLE 2 (Hull–White). Again assume $d=1$ but now take $\sigma_t(x) = \sigma\exp(-\alpha x)$, where $\sigma$ and $\alpha$ are positive constants. Straightforward integration imply

$$\begin{aligned}r_t(0) = r_0(t) &+ \frac{\sigma^2}{\alpha} \int_0^t e^{-\alpha(t-s)} \int_0^s [1 - e^{-\alpha(t-\theta)}]\phi_H(s-\theta)\, d\theta\, ds \\ &+ \frac{\sigma^2}{\alpha} \int_0^t [1 - e^{-\alpha(t-s)}] \int_0^s e^{-\alpha(t-\theta)}\phi_H(s-\theta)\, d\theta\, ds \\ &+ \sigma \int_0^t e^{-\alpha(t-s)}\, d\tilde{\beta}_s,\end{aligned}$$

which is consistent with the Hull–White model or the Vasicek model with time-varying drift parameters. Furthermore, in the fBm setting this model is $k$-arbitrage free for every $k>0$.

REMARK 3.9. One can easily extend the results of this section to a cylindrical fBm in the HJM equation (2.2) by imposing standard growth conditions on the volatility in (2.6), (2.7), (2.8) and (2.9). Moreover, the arguments used in Lemma 3.2 are also valid in this case due to the independence of infinitely many fBm.

**4. Consistency for fractional HJM models.** In this section, we study consistency problems related to HJM models introduced in the previous section. Without any loss of generality we now assume that the following dynamics takes place:

$$(4.1) \quad dr^{x_0}(t) = (Ar(t) + \mathcal{S}_H\sigma(t))\, dt + \sum_{j=1}^d \sigma^j d\beta^j(t), \qquad r_0 = x_0 \in E,$$

where $\mathcal{S}_H$ is given by (3.15). We study the important case of time-homogeneous HJM models, in the sense that the volatility $\sigma \in \mathcal{L}(U; E)$ does not depend on time $t \in [0, T]$ where $0 < T < \infty$ is a fixed terminal time.

We want to stress here that the main invariance result of this section (Theorem 4.2) holds for any $H \geq 1/2$ when dealing with general time-dependent drifts $t \mapsto \alpha_t \in E$. We take $\mathcal{S}_H \sigma(t)$ in (4.1) only because it is the canonical drift when calibrating HJM models from the observed forward curves. Moreover, it is straightforward the extension of Theorem 4.2 to the case of a cylindrical fBm and a Lipschitz vector field as the drift.

Let $\mathcal{P}$ be the interest rate model produced by $r_t$ and let $\mathcal{M}$ be a parameterized family of smooth forward curves (e.g., Nelson–Siegel or Svensson families). We recall that a pair $(\mathcal{P}, \mathcal{M})$ is *consistent* if all forward curves which may be produced by the interest rate model $\mathcal{P}$ are contained within the family $\mathcal{M}$, provided that the initial curve is in $\mathcal{M}$. There are several reasons, why in practice, one is interested in consistent pairs $(\mathcal{P}, \mathcal{M})$ with respect to HJM dynamics (see [3, 10]). In particular, the following questions are of great importance in calibrating interest rate models:

*Given an interest rate model $\mathcal{P}$ and a family of forward rate curves $\mathcal{M}$, what are the necessary and sufficient conditions for consistency? Let $\mathcal{M}$ be an exponential-polynomial family of smooth forward curves. Is there a nontrivial $\mathcal{P}$ which is consistent to $\mathcal{M}$?*

The remainder of this paper will be devoted to answer the above problems in the fBm setting.

4.1. *Consistent pairs $(\mathcal{P}, \mathcal{M})$.* In this section, we give a fairly complete characterization of a given $\mathcal{M}$ to be consistent with respect to $\mathcal{P}$. We adopt an abstract framework by considering $\mathcal{M}$ as a finite-dimensional smooth submanifold of $E$. The concept of invariance used in this work is the following:

DEFINITION 4.1. A closed set $K \subset E$ is said to be *invariant* for the forward rate $(r(t))_{0 \leq t \leq T}$ when

$$\mathbb{P}(r^{x_0}(t) \in K, \forall t \in [0, T]) = 1 \qquad \text{for every } x_0 \in K.$$

Hence the pair $(\mathcal{P}, \mathcal{M})$ is consistent if and only if $\mathcal{M}$ is invariant for the correspondent forward rate $r_t$. The main difficulty in characterizing consistent pairs $(\mathcal{P}, \mathcal{M})$ is the obtention of the topological support for the law of $\{r(t); 0 \leq t \leq T\}$ on the space $\mathcal{C}(0, T; E)$ of the $E$-valued continuous functions on $[0, T]$. We recall that the topological support $\operatorname{supp} \mu$ of a probability measure $\mu$ on a Polish space $\mathcal{X}$ is the smallest closed set in $\mathcal{X}$ with total mass.

Recall that the forward rate satisfies the following equation:

$$r(t) = S(t)x_0 + \int_0^t S(t-s)\mathcal{S}_H \sigma(s)\, ds + \sum_{i=1}^d \int_0^t S(t-s)\sigma^i\, d\beta^i(s),$$

where $\{S(t), t \geq 0\}$ is the right-shift semigroup acting on the Hilbert space $E$. Let us denote

$$
(4.2) \qquad Z(t) := \sum_{j=1}^{d} \int_0^t S(t-s) \sigma^j \, d\beta^j(s).
$$

Our main task is to characterize $\operatorname{supp} \mathbb{P}_Z$ on the space $\mathcal{C}(0, T; E)$. For this purpose, we take advantage of the fact that fBm is a centered Gaussian process. The theory of Gaussian processes provides a sharp characterization for the support of the measure $\mathbb{P}_Z$. A direct (but lengthy) calculation shows that the law of $Z(\cdot)$ in $L^2(0, T; E)$ is a symmetric Gaussian measure whose covariance operator is given by

$$
\Lambda_H \varphi(t) := \int_0^T g_H(t, s) \varphi(s) \, ds,
$$

where

$$
g_H(t, s) := \int_0^{s \wedge t} \int_0^{s \wedge t} S(t-v) \sigma \sigma^* S^*(s-u) \phi_H(u-v) \, du \, dv.
$$

By condition (2.7) it follows that $\mathbb{P}_Z$ is concentrated on $\mathcal{C}_0 = \{u \in \mathcal{C}([0, T]; E) : u(0) = 0\}$. Therefore, the closure of $\operatorname{Image} \Lambda_H^{1/2}$ in the $\mathcal{C}_0$-topology is the support of $\mathbb{P}_Z$. This fact would lead to a straightforward characterization of $\operatorname{supp} \mathbb{P}_Z$ as long as we are able to calculate the square root of the covariance operator $\Lambda_H$. In fact, a direct calculation proves to be quite difficult. Moreover, it is very hard (maybe unlikely) to find explicitly a bounded linear operator $\mathcal{A}$ such that $\Lambda_H = \mathcal{A}\mathcal{A}^*$; see Corollary B.4 in [7]. Therefore other nondirect techniques should be applied.

In the sequel, we consider the Wiener space $(\mathcal{W}, \mathbb{H}, \mathbf{P})$ of the $\mathbb{R}^d$-valued fBm, where $\mathcal{W}$ is the space of the $\mathbb{R}^d$-valued continuous functions $f$ on $[0, T]$ such that $f(0) = 0$, $\mathbb{H}$ is the correspondent Cameron–Martin space defined in (3.18) and $\mathbf{P}$ is the Wiener measure on $\mathcal{W}$. We have the following sufficient conditions for inclusions of the support of the law of an abstract Wiener functional $\mathcal{V} : \mathcal{W} \to \mathcal{X}$. See Aida, Kusuoka and Stroock [1] for the details.

PROPOSITION 4.1. *Let $\mathcal{V} : \mathcal{W} \to \mathcal{X}$ be a measurable map, where $\mathcal{X}$ is a separable Banach space.*

(i) *Let $\zeta_1 : \mathbb{H} \to \mathcal{X}$ be a measurable map, and let $\mathcal{J}_n : \mathcal{W} \to \mathbb{H}$ be a sequence of random elements such that for any $\varepsilon > 0$,*

$$
(4.3) \qquad \lim_n \mathbf{P}(\|\mathcal{V} - \zeta_1 \circ \mathcal{J}_n\|_{\mathcal{X}} > \varepsilon) = 0.
$$

*Then*

$$
\operatorname{supp} \mathbf{P}_{\mathcal{V}} \subset \overline{\zeta_1(\mathbb{H})}.
$$

(ii) *Let $\zeta_2 : \mathbb{H} \to \mathcal{X}$ be a map, and for each fixed $h \in \mathbb{H}$ let $\mathcal{T}_n^h : \mathcal{W} \to \mathcal{W}$ be a sequence of measurable transformations such that $\mathbf{P}_{\mathcal{T}_n^h} \ll \mathbf{P}$ for every $n$, and for any $\varepsilon > 0$,*

$$(4.4) \qquad \limsup_n \mathbf{P}(\|\mathcal{V} \circ \mathcal{T}_n^h - \zeta_2(h)\|_{\mathcal{X}} < \varepsilon) > 0.$$

*Then $\operatorname{supp} \mathbf{P}_{\mathcal{V}} \supset \overline{\zeta_2(\mathbb{H})}$.*

The remainder of this section will be devoted to the characterization of the topological support of the forward rate $r : \Omega \to \mathcal{C}(0, T; E)$ by using conditions (4.3) and (4.4). Clearly, the main step is the characterization of the support of the probability measure $\mathbb{P}_Z$. In the sequel, we write $\mathbf{E}$ to denote the expectation with respect to $\mathbf{P}$.

4.2. *Invariance for HJM models.* We now introduce a polygonal approximation for the fBm. Let us recall the Volterra representation of the fBm

$$(4.5) \qquad \beta(t) = \int_0^t K(t,s) \, dW(s); \qquad 0 \leq t \leq T,$$

where $W$ is the unique Wiener process that provides the integral representation (4.5) and $K(t,s)$ is the kernel defined in (3.19). See, for example, [26] for more details.

REMARK 4.1. From the above representation we notice that $W$ is adapted to the filtration generated by the fBm $\beta$ and both processes generate the same filtration.

Let $\Pi = 0 = t_0 < t_1 < \cdots < t_n = T$ be a partition of $[0, T]$ where $t_k := k\frac{T}{n}$ and $|\Pi| := \max_{0 \leq j \leq n-1}(t_{j+1} - t_j) = \frac{T}{n}$. Let us consider the following polygonal approximations

$$(4.6) \qquad \beta_\Pi(t) := \int_0^t K(t,s) \, dW_\Pi(s) = \int_0^t K(t,s) \dot{W}_\Pi(s) \, ds,$$

where

$$W_\Pi(t) := W(t_j) + \frac{W(t_{j+1}) - W(t_j)}{(t_{j+1} - t_j)}(t - t_j)$$

for $t_j \leq t \leq t_{j+1}$; $j = 0, 1, \ldots, n-1$.

One can check (see [18]) that for every $\gamma < 1 - H$ there exists a constant $C_{H,\gamma}$ independent of $\Pi$ such that

$$(4.7) \qquad \mathbf{E} \sup_{0 \leq t \leq T} |\beta_\Pi(t) - \beta(t)| \leq C_{H,\gamma} |\Pi|^\gamma.$$

If $\omega \in \mathcal{W}$ and $|\Pi| = T/n$ then we define $\omega^{(n)}(t) = (\omega_1^{(n)}(t), \ldots, \omega_d^{(n)}(t))$ where

$$\omega_i^{(n)}(t) := \int_0^t K(t,s) \dot{W}_{\Pi,i}(\omega)(s)\, ds, \qquad 1 \leq i \leq d.$$

Obviously $\omega^{(n)} \in \mathbb{H}$ for all $n \geq 1$ and $\omega \in \mathcal{W}$. For each $h \in \mathbb{H}$ we define

(4.8) $$\mathcal{T}_n^h \omega := \omega + (h - \omega^{(n)}).$$

The next result is a straightforward application of a Girsanov theorem [18] in the fBm setting.

LEMMA 4.1. *If $h \in \mathbb{H}$ then $\mathbf{P}_{\mathcal{T}_n^h} \ll \mathbf{P}$ for all $n \geq 1$.*

It will be convenient in the sequel to make use of fractional integration and differentiation. For $\alpha > 0$, we define the fractional integration operator $I^\alpha$ and the correspondent fractional differentiation operator $D^\alpha$ by

(4.9) $$I^\alpha f(t) = \frac{1}{\Gamma(\alpha)} \int_0^t (t-s)^{\alpha-1} f(s)\, ds;$$

(4.10) $$D^\alpha f(t) = \frac{1}{1 - \Gamma(\alpha)} \frac{d}{dt} \int_0^t (t-s)^{-\alpha} f(s)\, ds.$$

For a comprehensive survey of the properties of these operators, see [25]. The most important property that we are going to use here is that $I^\alpha$ and $D^\alpha$ are each other's inverses. The following result is crucial to get (4.3) in Proposition 4.1. In the sequel, we write $(\Psi \cdot \beta)$ and $(\Psi \cdot \beta_\Pi)$ to denote the Paley–Wiener integrals with respect to $\beta$ and $\beta_\Pi$, respectively.

PROPOSITION 4.2. *Let $\beta_\Pi$ be the polygonal approximation of the real-valued fBm. If $\Psi \in L^2(0,T;E)$ then*

$$\lim_{\|\Pi\| \to 0} \mathbf{E} \sup_{0 \leq t \leq T} \|(\Psi \cdot \beta)(t) - (\Psi \cdot \beta_\Pi)(t)\|_E = 0.$$

PROOF. We proceed by approximating $\Psi$ by step functions $f$. Assume that

$$f(s) = \sum_{i=0}^{n-1} \alpha_i \chi_{[s_i, s_{i+1})}(s); \qquad 0 = s_0 < s_1 < \cdots < s_n = T.$$

By the semigroup property of fractional integrals and taking into account that $D^{H+1/2}$ is the inverse $I^{H+1/2}$ it follows that

$$\|(f \cdot \beta)(t) - (f \cdot \beta_\Pi)(t)\|_E$$

$$= \left\| \sum_{i=0}^{n-1} \alpha_i \left[ (\beta(t_{i+1} \wedge t) - \beta(t_i \wedge t)) - \int_{t_i \wedge t}^{t_{i+1} \wedge t} \theta_H \beta_\Pi(s) \, ds \right] \right\|_E$$

$$\leq \sum_{i=0}^{n-1} \|\alpha_i\|_E |(\beta(t_{i+1} \wedge t) - \beta(t_i \wedge t)) - (\beta_\Pi(t_{i+1} \wedge t) - \beta_\Pi(t_i \wedge t))|.$$

By the estimate (4.7) we conclude that the assertion is true for step functions. Now let us consider $\Psi \in L^2(0, T; E)$ and a sequence $(f_n)_{n \geq 1}$ of step functions which converges to $\Psi$ in $L^2(0, T; E)$. We have

$$\sup_{0 \leq t \leq T} \|(\Psi \cdot \beta)(t) - (\Psi \cdot \beta_\Pi)(t)\|_E \leq \sup_{0 \leq t \leq T} \|(\Psi \cdot \beta)(t) - (f_n \cdot \beta)(t)\|_E$$
$$+ \sup_{0 \leq t \leq T} \|(f_n \cdot \beta)(t) - (f_n \cdot \beta_\Pi)(t)\|_E$$
$$+ \sup_{0 \leq t \leq T} \|(f_n \cdot \beta_\Pi)(t) - (\Psi \cdot \beta_\Pi)(t)\|_E$$
$$= T_1(n) + T_2(n, \Pi) + T_3(n, \Pi).$$

By the first step we only need to estimate $T_1$ and $T_3$. Hölder's inequality yields

$$(4.11) \quad T_3(n, \Pi) \leq \|f_n - \Psi\|_{L^2(0,T;E)} \|\theta_H \beta_\Pi\|_{L^2(0,T;\mathbb{R})} < \infty \quad \text{a.s.,}$$

where $\theta_H := I^{H-1/2} \circ D^{H+1/2}$ and $\|\theta_H \beta_\Pi\|_{L^2(0,T;\mathbb{R})}$ is square integrable. Therefore we can conclude that $\lim_{n \to \infty} \mathbf{E} T_3(n, \Pi) = 0$ for each partition $\Pi$. It remains to estimate $T_1$. For this we shall use the factorization method on the fractional Wiener integral. Recall the identity

$$(4.12) \quad \frac{\pi}{\sin \pi \alpha} = \int_\sigma^t (t-s)^{\alpha-1} (s-\sigma)^{-\alpha} \, ds; \quad \sigma \leq s \leq t, \ 0 < \alpha < 1.$$

Fix $0 < \alpha < 1/2$ and $p > 1/2\alpha$. By using (4.12) and the stochastic Fubini theorem for fractional Wiener integrals [20], we may write

$$((\Psi - f_n) \cdot \beta)(t) = \frac{\sin \pi \alpha}{\pi} \int_0^t (t-s)^{\alpha-1} y_m(s) \, ds,$$

where $y_m(s) := \int_0^s (\Psi - f_m)(\sigma)(s - \sigma)^{-\alpha} \, d\beta(\sigma)$. Hölder's inequality yields

$$\sup_{0 \leq t \leq T} \|((\Psi - f_n) \cdot \beta)(t)\|_E^{2p} \leq C_1 \int_0^T \|y_m(s)\|_E^{2p} \, ds,$$

where the constant $C_1$ depends only on $p, \alpha$ and $T$. We now choose $p = 1$. The ordinary Fubini theorem and the isometry of the fractional Wiener integral with the reproducing kernel Hilbert space of the fBm (see [2, 9])

allow us to write

$$\mathbf{E}T_1^2(n) \leq C_1 \int_0^T \mathbf{E}\|y_m(s)\|_E^2\, ds$$

$$= C_1 \int_0^T \int_0^s \int_0^s \langle(\Psi - f_n)(u)(s-u)^{-\alpha}, (\Psi - f_n)(v)(s-v)^{-\alpha}\rangle_E$$
$$\times \phi_H(u-v)\, du\, dv\, ds.$$

By the estimate (11) in [2] we can find positive constants $C_2$ and $C_3$ such that

$$\mathbf{E}T_1^2(n) \leq C_2 \int_0^T \int_0^s \|(\Psi - f_n)(u)(s-u)^{-\alpha}\|_E^2\, du\, ds$$
$$\leq C_3 \|\Psi - f_n\|_{L^2(0,T;E)}.$$

Summing up all the estimates we complete the proof of the proposition. □

We say that a closed set $K$ in $E$ is invariant for the evolution equation

(4.13)
$$\frac{d}{dt}y^{(x_0,u)}(t) = Ay^{(x_0,u)}(t) + \mathcal{S}_H\sigma(t) + \sigma I^{H-1/2}u(t),$$
$$y(0) = x_0 \in E,$$

if for each initial condition $x_0 \in K$ and a control $u \in L^2([0,T];U)$ we have

$$y^{(x_0,u)}(t) \in K \quad \text{for all } t \in [0,T].$$

In accordance with Proposition 4.1, we are now in position to define the following mappings:

(4.14) $$\zeta_1 h(t) := \sum_{i=1}^d \int_0^t S(t-s)\sigma^i \theta_H h_i(s)\, ds, \quad h \in \mathbb{H},$$

(4.15) $$\mathcal{J}_n(\omega) := \omega^{(n)}, \quad \omega \in \mathcal{W},$$

where $\theta_H = I^{H-1/2} \circ D^{H+1/2}$.

THEOREM 4.1. *A closed set is invariant for the differential equation (4.13) if and only if it is invariant for the HJM equation (4.1). In particular,*

$$\operatorname{supp} \mathbb{P}_{r^{x_0}} = \overline{\{y^{(x_0,u)}; u \in L^2(0,T;U)\}}.$$

PROOF. We apply Proposition 4.1 to the Wiener functional $Z$ defined in (4.2) with the correspondent mappings $\zeta_1$, $\mathcal{J}_n$ and $\mathcal{T}_n^h$, defined in (4.14), (4.15) and (4.8), respectively. Conditions (4.3) and (4.4) in Proposition 4.1 are direct consequences of Proposition 4.2 and Lemma 4.1. We then have the following characterization

$$\operatorname{supp} Z = \zeta_1 \overline{(I^{H+1/2}(L^2(0,T;U)))}$$

and therefore the law of $Z$ is concentrated on the set of continuous functions of the form

$$\int_0^t S(t-s)\sigma I^{H-1/2} h(s)\,ds, \qquad h \in L^2([0,T];U).$$

Now the proof follows the same lines of ([22], Proposition 1.1), and therefore we omit the details. □

4.3. *Nagumo conditions and finite-dimensional invariant manifolds.* In this section we prove the main result of this section. If $\mathcal{M}$ is a $C^1$-manifold in $E$ then we write $T_x\mathcal{M}$ the associated tangent space at $x \in \mathcal{M}$. We now provide Nagumo-type conditions for an HJM model to be invariant with respect to a given smooth manifold.

THEOREM 4.2. *Let $\mathcal{M}$ be a $C^1$-submanifold in $E$, closed as a set and $\mathcal{M} \subset \operatorname{Dom}(A)$. Then $\mathcal{M}$ is invariant for the stochastic equation (4.1) if and only if*

$$Ax + \mathcal{S}_H \sigma(t) + \sigma \nu \in T_x \mathcal{M} \tag{4.16}$$

*for each $x \in \mathcal{M}$, $t \in [0,T]$ and $\nu \in U$.*

PROOF. Let $\mathcal{E}$ be the set of $U$-valued piecewise constant functions. We claim that a closed set $K$ is invariant for (4.13) if and only if its mild solution satisfies the following condition: for each $x \in K$ and $v \in \mathcal{E}$ we have $y^{(x,v)}(t) \in K$ for all $t \in [0,T]$. We fix an arbitrary $u \in L^2(0,T;U)$ and let us consider a sequence of step functions $u_n$ converging to $u$ in $L^2(0,T;U)$. Then

$$(4.17)\quad \|y^{(x,u_n)}(t) - y^{(x,u)}(t)\|_E \leq \sup_{0 \leq t \leq T} \left\| \int_0^t S(t-s)\sigma I^{H-1/2}(u_n - u)(s)\,ds \right\|_E.$$

By Hölder inequality we have

$$\begin{aligned}
&\sup_{0 \leq t \leq T} \left\| \int_0^t S(t-s)\sigma I^{H-1/2}(u_n - u)(s)\,ds \right\|_E \\
&\qquad \leq C \left( \int_0^T \|\sigma(u_n - u)(r)\|_E^2\,dr \right)^{1/2},
\end{aligned} \tag{4.18}$$

where $C$ is a positive constant which depends on $T$ and $H$. Since $\sigma$ is bounded we then have inequalities (4.17) and (4.18) imply that a closed set $K$ is invariant for (4.13) if and only if $y^{(x,v)}(t) \in K$, $t \in [0,T]$ for all $x \in K$ and all piecewise constant $U$-valued function $v$. Thus proving our first claim.

Now let $\mathcal{M} \subset E$ be a closed $C^1$-submanifold where $\mathcal{M} \subset \text{Dom}(A)$. By Theorem 4.1, Theorem 2 in [19], and the linearity of $U$, $I^{H-1/2}$ and $\sigma$ we may conclude relation (4.16) $\square$

We are now in position to characterize a given finite-dimensional invariant submanifold. In the sequel, we write $\text{Im}\,\sigma := \sigma U$.

COROLLARY 4.1. *Let $\mathcal{M}$ be a finite-dimensional $C^1$-submanifold in $E$ (closed as a set). Then $\mathcal{M}$ is invariant for an HJM model given by (4.1) if and only if $\mathcal{M} \subset \text{Dom}(A)$, and*

$$Ax \in T_x\mathcal{M}, \tag{4.19}$$

$$\mathcal{S}_H\sigma(t) + \text{Im}\,\sigma \subset T_x\mathcal{M} \tag{4.20}$$

*for every $t \in [0,T]$ and $x \in \mathcal{M}$.*

PROOF. Assume that $\mathcal{M}$ is invariant for an HJM model given by (4.1). We may apply the same arguments as in Lemma 2.3 and 2.4 in [22] to show that $\mathcal{M} \subset \text{Dom}(A)$ and therefore

$$\mathbb{P}(r^{x_0}(t) \in \text{Dom}(A), \forall t \in [0,T]) = 1 \qquad \text{for every } x_0 \in \mathcal{M}.$$

From Theorem 4.1 we know that every mild solution of (4.13) is also a strong solution such that $\mathcal{S}_H\sigma(0) = \sigma I^{H-1/2}u(0) = 0$ for every $u \in L^2(0,T;U)$. By differentiating at $t = 0$ and observing (4.16), we conclude relations (4.19) and (4.20). Conversely, let $x \in \mathcal{M}$, $v \in U$ and $t \in [0,T]$. By Theorem 4.2 it is sufficient to check (4.16). But this is a straightforward calculation using the parameterizations in $\mathcal{M}$. $\square$

**5. Nelson–Siegel family.** In this section, we investigate the Nelson–Siegel exponential family $\mathcal{M} = \{F(\cdot,y); y \in \mathcal{Y}\}$ widely used to fit term structure of interest rates. The form of the curve is given by the following expression

$$F(x,y) = y_1 + y_2 e^{-y_4 x} + y_3 x e^{-y_4 x}, \qquad x \geq 0, \tag{5.1}$$

where we restrict the parameters to the following state space $\mathcal{Y} := \{y = (y_1, \ldots, y_4) \in \mathbb{R}^4 | y_4 \neq 0\}$.

As an example, we now study simple interest rate models. By using Corollary 4.1 [in particular relations (4.19) and (4.20)], the calculations are minor modifications from [3] so we just give a sketch of the proofs.

5.1. *Ho–Lee model.* Let us consider the one-factor model under the Ho–Lee volatility structure given by a constant volatility, that is, $\sigma_t(x) = \sigma$, for all $(t,x) \in \mathbb{R}_+^2$. In this case, the drift restriction is given by

$$\mathcal{S}_H \sigma(t,x) = \sigma^2 (\varrho_H^1(t,x) + \varrho_H^2(t)), \qquad (t,x) \in \mathbb{R}_+^2; \tag{5.2}$$

where $\varrho_H^1(t,x) = 2x \int_0^t \phi_H(t-\theta)\,d\theta$ and $\varrho_H^2(t) = t \int_0^t \phi_H(t-\theta)\,d\theta - \int_0^t \theta \phi_H(t-\theta)\,d\theta$. One should notice that relation (4.19) is satisfied but because of the term $\varrho_H^1(t,x)$ in (5.2), relation (4.20) is not possible and therefore by Corollary 4.1 we conclude that the Ho–Lee model is not consistent with the Nelson–Siegel family.

5.2. *Hull–White model.* Let us consider the one-factor model under the Hull–White volatility structure given by $\sigma_t(x) = \sigma e^{-\alpha x}$ where $\alpha$ and $\sigma$ are positive constants. In this case, the drift restriction is given by

$$\begin{aligned}\mathcal{S}_H \sigma(t,x) &= \frac{\sigma^2}{\alpha} e^{-\alpha x} \left[ \int_0^t (1 + e^{-\alpha(t-\theta)}) \phi_H(t-\theta)\,d\theta \right] \\ &\quad - \frac{2\sigma^2}{\alpha} e^{-2\alpha x} \int_0^t e^{-\alpha(t-\theta)} \phi_H(t-\theta)\,d\theta \end{aligned} \tag{5.3}$$

for each $(t,x) \in \mathbb{R}_+^2$. By considering the full state space $\mathcal{Y}$, clearly the Hull–White model cannot be consistent with the Nelson–Siegel family. By restricting the state space to $\mathcal{Y}^\alpha = \{y = (y_1,\ldots,y_4) | y_4 = \alpha\}$, the curve shape is then given by

$$F(x,y) = y_1 + y_2 e^{-\alpha x} + y_3 x e^{-\alpha x}.$$

Due to the second term in (5.3), the fractional Hull–White model is not consistent with the Nelson–Siegel family on $\mathcal{Y}^\alpha$. In fact, the following result is not surprising in view of the previous examples.

PROPOSITION 5.1. *There is no nontrivial fractional interest rate model with deterministic volatility which is consistent with the Nelson–Siegel family.*

PROOF. By using Theorem 4.1 the proof is analogous to [3], Proposition 7.1. □

REMARK 5.1. One can easily find consistent families by modifying, for example, the Nelson–Siegel family. By using relations (4.19) and (4.20), the arguments are completely analogous to [3].

## APPENDIX: INTEGRATION FOR $\mathcal{C}_{\mathbb{R}_+}$-VALUED PROCESSES

In this section, we give the details of the integral introduced in Section 3 and we keep the same notation introduced there. From Section 3 we know that if $\mu \in V$ and $G$ satisfies (H1) then $\int_0^\cdot \mu_s \, dG_s$ is well defined. In the sequel, $t_i^n := iT^*/2^n$ for $i = 0, 1, \ldots, 2^n$; $n \geq 1$ and $0 < T^* < \infty$.

LEMMA A.1. *If $\mu \in V$ and $G$ satisfies assumption* (H1) *then*

$$\lim_{n \to \infty} \mathbb{E} \left\| \sum_{i=0}^{2^n - 1} \mu_{t_i^n}(G_{t_{i+1}^n \wedge \cdot} - G_{t_i^n \wedge \cdot}) - \int_0^\cdot \mu_s \, dG_s \right\|_\infty = 0.$$

PROOF. Straightforward estimates by approximating $\mu$ by simple process. □

Let $\overline{G}_n$ and $\underline{G}_n$ be the upper and the lower approximations of $G$, respectively, along the partition $(t_i^n)$. Let $\int_0^t \overline{G}_n(s) \, d\mu_s$ and $\int_0^t \underline{G}_n(s) \, d\mu_s$ be the respective upper and lower integrals.

LEMMA A.2. *Assume that $\mu \in V$ where* (H2) *holds and consider $G$ a stochastic process such that* (H1) *holds. Then:*

(a) $\lim_{n \to \infty} \mathbb{E} \sup_{0 \leq t \leq T^*} |\int_0^t \overline{G}_n \, d\mu - \int_0^t \underline{G}_n \, d\mu| = 0$,
(b) $\lim_{n, m \to \infty} \mathbb{E} \sup_{0 \leq t \leq T^*} |\int_0^t \overline{G}_n \, d\mu - \int_0^t \overline{G}_m \, d\mu| = 0$.

PROOF. By the continuity of $G$ and assumptions (H1)–(H2) we may apply dominated convergence theorem to conclude part (a). Again by continuity, it follows that $\sup_{0 \leq s, T \leq T^*} |\overline{G}_n(s; T) - \overline{G}_m(s; T)| \to 0$ a.s. as $n, m \to \infty$. Moreover, it is bounded by $C \sup_{s, T \in [0, T^*]^2} |G(s, T)|$. Again, assumptions (H1)–(H2) and dominated convergence theorem allow us to conclude part (b). □

By Lemma A.2 we shall define

$$\int_0^t G_s \, d\mu_s := \lim_{n \to \infty} \int_0^t \overline{G}_n(s) \, d\mu_s = \lim_{n \to \infty} \int_0^t \underline{G}_n(s) \, d\mu_s.$$

The next result is a straightforward integration by part formula.

PROPOSITION A.1. *Assume that assumptions* (H1) *and* (H2) *hold. Then*

(A.1) $$\int_0^{T^*} G_s \, d\mu_s + \int_0^{T^*} \mu_s \, dG_s = G_{T^*} \mu_{T^*} - G_0 \mu_0.$$

PROOF. By writing a telescoping sum we have

$$\sum_{i=0}^{2^n-1} (G_{t_{i+1}^n} - G_{t_i^n})(\mu_{t_{i+1}^n} - \mu_{t_i^n})$$
$$= G_{T^*}\mu_{T^*} - G_0\mu_0$$
$$- \sum_{i=0}^{2^n-1} (G_{t_{i+1}^n} - G_{t_i^n})\mu_{t_i^n} - \sum_{i=0}^{2^n-1} (\mu_{t_{i+1}^n} - \mu_{t_i^n})G_{t_i^n},$$

a.s. for all $n \geq 1$. By Lemmas A.1 and A.2 we only need to show that the left side goes to zero as $n \to \infty$. But this is an immediate consequence of hypotheses (H1)–(H2) together with the continuity of $G$. $\square$

**Acknowledgments.** The author is particularly grateful to Pedro Catuogno for many fruitful discussions and suggestions. The author would also like to thank David Elworthy for his kind hospitality during the visit to the Mathematics Research Centre at the University of Warwick.

## REFERENCES

[1] AIDA, S., KUSUOKA, S. and STROOCK, D. (1993). On the support of Wiener functionals. In *Asymptotic Problems in Probability Theory: Wiener Functionals and Asymptotics (Sanda/Kyoto, 1990). Pitman Res. Notes Math. Ser.* **294** 3–34. Longman, Harlow. MR1354161

[2] ALÒS, E. and NUALART, D. (2003). Stochastic integration with respect to the fractional Brownian motion. *Stochastics Rep.* **75** 129–152. MR1978896

[3] BJÖRK, T. and CHRISTENSEN, B. J. (1999). Interest rate dynamics and consistent forward rate curves. *Math. Finance* **9** 323–348. MR1849252

[4] BJÖRK, T., DI MASI, G., KABANOV, W. and RUNGGALDIER, Y. (1997). Towards a general theory of bond markets. *Finance Stoch.* **1** 141–174.

[5] BJÖRK, T. and HULT, H. (2005). A note on Wick products and the fractional Black–Scholes model. *Finance Stoch.* **9** 197–209. MR2211124

[6] CARMONA, R. and TEHRANCHI, M. (2004). A characterization of hedging portfolios for interest rate contingent claims. *Ann. Appl. Probab.* **14** 1267–1294. MR2071423

[7] DA PRATO, G. and ZABCZYK, J. (1992). *Stochastic Equations in Infinite Dimensions. Encyclopedia of Mathematics and Its Applications* **44**. Cambridge Univ. Press, Cambridge. MR1207136

[8] DE DONNO, M. and PRATELLI, M. (2007). A theory of stochastic integration for bond markets. *Ann. Appl. Probab.* **15** 2773–2791.

[9] DUNCAN, T. E., MASLOWSKI, B. and PASIK-DUNCAN, B. (2002). Fractional Brownian motion and stochastic equations in Hilbert spaces. *Stoch. Dyn.* **2** 225–250. MR1912142

[10] FILIPOVIĆ, D. (2001). *Consistency Problems for Heath–Jarrow–Morton Interest Rate Models. Lecture Notes in Mathematics* **1760**. Springer, Berlin. MR1828523

[11] FILIPOVIĆ, D. and TEICHMANN, J. (2003). Existence of invariant manifolds for stochastic equations in infinite dimension. *J. Funct. Anal.* **197** 398–432. MR1960419

[12] GAPEEV, P. V. (2004). On arbitrage and Markovian short rates in fractional bond markets. *Statist. Probab. Lett.* **70** 211–222. MR2108087
[13] GRIPENBERG, G. and NORROS, I. (1996). On the prediction of fractional Brownian motion. *J. Appl. Probab.* **33** 400–410. MR1385349
[14] GUASONI, P. (2006). No arbitrage under transaction costs, with fractional Brownian motion and beyond. *Math. Finance* **16** 569–582. MR2239592
[15] GUASONI, P., RÁSONYI, M. and SCHACHERMAYER, W. (2008). The fundamental theorem of asset pricing for continuous process under small transaction costs. *Annals of Finance*. To appear.
[16] HAIRER, M. and OHASHI, A. (2007). Ergodic theory for SDEs with extrinsic memory. *Ann. Probab.* **35** 1950–1977. MR2349580
[17] HEATH, D., JARROW, R. and MORTON, A. (1992). Bond pricing and the term structure of interest rates: A new metodology for contingent claims valuation. *Econometrica* **60** 77–105.
[18] HU, Y. (2005). Integral transformations and anticipative calculus for fractional Brownian motions. *Mem. Amer. Math. Soc.* **175** 127. MR2130224
[19] JASHIMIAK, W. (1997). A note on invariance for semilinear differential equations. *Bull. Polish Acad. Sci. Math.* **45** 181–185. MR1466843
[20] KRVAVICH, Y. V. and MĪSHURA, Y. S. (2001). Differentiability of fractional integrals whose kernels contain fractional Brownian motions. *Ukraïn. Mat. Zh.* **53** 35–47. MR1834637
[21] MCCARTHY, J., DISARIO, R., SARAOGLU, H. and LI, H. (2004). Tests of long-range dependence of interest rates using wavelets. *Quarterly R. Economics. Finance* **44** 180–189.
[22] NAKAYAMA, T. (2004). Viability theorem for SPDE's including HJM framework. *J. Math. Sci. Univ. Tokyo* **11** 313–324. MR2097528
[23] NORROS, I., VALKEILA, E. and VIRTAMO, J. (1999). An elementary approach to a Girsanov formula and other analytical results on fractional Brownian motions. *Bernoulli* **5** 571–587. MR1704556
[24] RUSSO, F. and VALLOIS, P. (2000). Stochastic calculus with respect to continuous finite quadratic variation processes. *Stochastics Rep.* **70** 1–40. MR1785063
[25] SAMKO, S. G., KILBAS, A. A. and MARICHEV, O. I. (1993). *Fractional Integrals and Derivatives: Theory and Applications*. Gordon and Breach, Yverdon. MR1347689
[26] SAMORODNITSKY, G. and TAQQU, M. S. (1994). *Stable Non-Gaussian Random Processes: Stochastic Models with Infinite Variance*. Chapman & Hall, New York. MR1280932
[27] SOTTINEN, T. and VALKEILA, E. (2001). Fractional Brownian motion as a model in finance. Preprint 302, Dept. Mathematics, Univ. Helsinki.

IBMEC BUSINESS SCHOOL–SÃO PAULO
04546-042-SÃO PAULO, SP
BRAZIL
AND
DEPARTAMENTO DE MATEMÁTICA
UNIVERSIDADE ESTADUAL DE CAMPINAS
UNICAMP
13.083-859-CAMPINAS, SP
BRAZIL
E-MAIL: ohashi@ime.unicamp.br
        AlbertoMFO@isp.edu.br